\documentstyle[preprint,floats,aps,psfig,epsfig]{revtex}

\begin{document}

\title{Instability of the Fermi-liquid fixed point in an extended Kondo model}

\author{M. Lavagna$^{1,*}$, A. Jerez$^{2}$ and D. Bensimon$^{1,3}$}
\address{$^1$Commissariat \`a l'Energie Atomique, DRFMC/SPSMS,
17, rue des Martyrs, 38054 Grenoble Cedex 9, France}
\address{$^2$European Synchrotron Radiation Facility, 6, rue Jules Horowitz,
38043 Grenoble Cedex 9, France}
\address{$^3$Department of Appled Physics, Hongo 7-3-1, University of Tokyo,  
Tokyo 113-8656, Japan}

\maketitle 

\begin{abstract}
We study an extended SU(N) single-impurity Kondo model in
which the impurity spin is described by a combination of 
Abrikosov fermions and Schwinger bosons. Our aim is to describe
both the quasiparticle-like excitations and the locally critical
modes observed in various physical situations, including
non-Fermi liquid (NFL) behavior in heavy fermions in the
vicinity of a quantum critical point and anomalous transport
properties in quantum wires. In contrast with models with 
either pure bosonic or pure fermionic impurities, the strong 
coupling fixed point is unstable against the conduction
electron kinetic term under certain conditions. The stability 
region of the strong coupling fixed point coincides with the 
region where the partially screened, effective impurity repels 
the electrons on adjacent sites. In the instability region, the 
impurity tends to attract $(N-1)$ electrons to the neighboring 
sites, giving rise to a double-stage Kondo effect with additional 
screening of the impurity.
\end{abstract}

\bigskip

\section{Introduction}

One of the most striking properties discovered in Heavy-Fermion systems these
last years is the existence of a non-Fermi-liquid behavior 
\cite{vl96,steglich96,mathur98}
in the disordered 
phase close to the magnetic quantum critical point with a temperature dependence 
of the physical quantities which differs from that of a standard Fermi-liquid. 
Recent results obtained in $CeCu_{5.9}Au_{0.1}$ by inelastic neutron scattering 
experiments \cite{schroder98,schroder00}
have shown the existence of anomalous $\omega /T$ scaling law for 
the dynamical spin susceptibility at the antiferromagnetic wavevector
which persists over the entire Brillouin zone. This indicates that 
the spin dynamics are critical not only at large length scales as 
in the itinerant magnetism picture \cite{hertz76,millis93}
but also at atomic length scales. 
It strongly suggests the presence of locally critical modes beyond 
the standard spin-fluctuation theory. Alternative theories have then been proposed 
to describe such a local quantum critical point. In this direction, 
we will mention recent calculations \cite{si01} based on dynamical mean field
theory 
(DMFT) which seem to lead to encouraging results such as the prediction of a scaling law 
for the dynamical susceptibility. 
The other approach which has been 
put forward to describe the local QCP is based on supersymmetric 
theory \cite{gca92,pl99,cpt1,cpt2}
in which the spin is described in a mixed fermionic-bosonic representation. 
The interest of the supersymmetric approach is to describe the quasiparticles 
and the local moments on an equal footing through the fermionic and the bosonic 
part of the spin respectively. It appears to be specially well-indicated in the 
case of the locally critical scenario in which the magnetic temperature scale 
$T_{N}$, and the Fermi scale $T_{K}$ (the Kondo temperature) below which the 
quasiparticles die, vanish at the same point $\delta _{C}$.

Let us now emphasize another aspect in the discussion of the 
breakdown of the Fermi-liquid theory. It has to do with the question of 
the instability of the strong coupling (SC) fixed point. A stable SC fixed 
point is usually associated with a local Fermi-liquid behavior. On the 
contrary, an instability of the SC fixed point is an indication for the
existence of an intermediate coupling fixed point with non-Fermi-liquid
behavior. The traditional source of instability of the SC
fixed point in the Kondo model is the presence of several channels for the 
conduction electrons with the existence of two regimes, the underscreened 
and the overscreened ones with very different associated behavior. 
In the one-channel antiferromagnetic single-impurity Kondo model, 
it is well known from 
renormalization group arguments, that the system flows to a stable
SC fixed point \cite{pwa67}
with a behavior of the system identified with
that of a local Fermi liquid \cite{pn76}. The situation is rather 
different in the 
case of the multichannel Kondo
model with a number $K$ of channels for the conduction electrons ($K>1$)
\cite{nb}. 
In the underscreened regime when $K<2S$ (where $S$ is the value of the 
spin in $SU(2)$), the
SC fixed point is stable. It becomes unstable in the other regime $K>2S$
named as the overscreened regime. The underscreened regime 
corresponds to the one-stage Kondo effect with the formation of an effective
spin $(S-1/2)$ resulting from the screening of the impurity spin by the
conduction electrons located on the same site. The system described by the
SC fixed point behaves as a local Fermi liquid. The
instability of the SC fixed point in the overscreened 
regime is associated with a multi-stage Kondo effect in which successively
the impurity spin is screened by conduction electrons on the same site, and
then the dressed impurity is screened by conduction electrons on the
neighboring site and so forth.
The overscreened regime leads to the existence of an intermediate
coupling fixed point with non-Fermi-liquid excitation spectrum and an anomalous residual 
entropy at zero temperature. 
It has been recently put forward \cite{cpt1,cpt2} that other sources of instability 
of the SC
fixed point may exist else but the multiplicity of the conduction electron
channels. Recent works have shown that the presence of a more general Kondo
impurity where the spin symmetry is extended from $SU(2)$ to $SU(N)$, 
and the impurity has mixed symmetry,
may also lead to an
unstability of the SC fixed point already in the one-channel case. In the
large $N$ limit, Coleman et al. \cite{cpt1} have found that the SC fixed point becomes
unstable as soon as the impurity parameter $q$ (defined below)
is larger than $N/2$ whatever the value of $2S$  (defined below)
is, giving rise to a
two-stage Kondo effect. This result opens the route for the existence of an
intermediate coupling fixed point with eventually non-Fermi-liquid
behavior. 

It is worth noticing that the supersymmetry theory, or more specifically 
the taking into account of general Kondo impurities
appears to offer valuable insights into the two issues raised by the
breakdown of the Fermi liquid theory that we have summarized above, i.e.
both the existence of locally critical modes and the question of the
instability of the SC fixed point. Somehow it seems that the consideration
of general Kondo impurities captures the physics present in real systems
with the coexistence of the screening of the spin by conduction electrons
responsible for the formation of quasiparticles, and the formation of
localized magnetic moment that persists and eventually leads to a phase
transition as the coupling to other impurities becomes dominant. 

The aim of this work is to study the $SU(N)$, extended
single-impurity Kondo model in the one-channel case \cite{ljb}. We would like to
investigate how the system behaves when not only the values of the
parameters $(2S,q)$ of the representation vary, but also the number $n_{d}$
of conduction electrons on the neighboring site does. We want to discuss
the effect of $n_{d}$ on the stability of the SC fixed point and to further
understand the nature of the screening  with the possibility of achieving
either a one-stage, two-stage or multi-stage Kondo effect depending on the
regime considered. Implications for the behavior of physical 
quantities will be given.

\section{Extended $SU(N)$ single-impurity Kondo model}

We consider a generalized, single-impurity, Kondo model with one channel of
conduction electrons and a spin symmetry group extended from $SU(2)$ to 
$SU(N)$. An impurity spin, $\mathbf{S}$, is placed at the origin (site $0$).
We deal with impurities that can be
realized by a combination of $2S$ bosonic ($b^\dagger_\alpha$) and 
$q-1$ fermionic ($f^\dagger_\alpha$) operators. 
The hamiltonian describing the model is written as 
\begin{equation}
H=\sum_{\mathbf{k},\alpha }\varepsilon _{\mathbf{k}}c_{\mathbf{k},\alpha
}^{\dagger }c_{\mathbf{k},\alpha }+J\sum_{A}\mathbf{S}^{A}\sum_{\alpha
,\beta }c_{\alpha }^{\dagger }(0) {\tau}_{\alpha \beta
}^{A}c_{\beta }(0)~,  \label{ham1}
\end{equation}
where $c_{\mathbf{k},\alpha }^{\dagger }$ is the creation operator of a
conduction electron with momentum $\mathbf{k}$, and $SU(N)$ spin index 
$\alpha =a,b,...,r_N$, $c_{\alpha }^{\dagger }(0)=\frac{1}{\sqrt{N_{S}}}
\sum_{\mathbf{k}}c_{\mathbf{k},\alpha }^{\dagger }$ 
, where $N_{S}$ is the number of sites, and 
${\tau }_{\alpha\beta}^{A}$ 
($A=1,\dots,N^{2}-1$) are the generators of the 
$SU(N)$ group
in the fundamental representation,
with $Tr[{\tau}^{A}{\tau}^{B}]=\delta _{AB}/2$. The
conduction electrons interact with the impurity spin $\mathbf{S}^{A}$ 
($A=1,\dots,N^{2}-1$), placed at the origin, via Kondo coupling, $J>0$. 
When the
impurity is in the fundamental representation, we recover the
Coqblin-Schrieffer model \cite{cs,hew} describing conduction electrons 
in interaction with
an impurity spin of angular momentum $j$, 
($N=2j+1$, $a=j,~b=j-1,\dots,r_N=-j$), resulting from the
combined spin and orbit exchange scattering.

\subsection{Strong-coupling fixed point}

We consider the case where $J\rightarrow \infty$ and we can
neglect the kinetic energy in Eq. \ref{ham1}. In this limit the model
can be solved exactly in terms of the invariants associated to the
spin of the electrons and of the impurities 
\cite{jaz}. The eigenvalues are of the
form
\begin{eqnarray}
E_{SC} = \frac{J}{2} \left[ {\cal C}_2(R_{SC})-{\cal C}_2(I)-{\cal C}_2(Y)
\right]
\label{ener}
\end{eqnarray}
where $I$ denotes the impurity spin, $Y$ the spin of the $n_c$ 
conduction electrons coupled to the impurity 
at the origin, and $R_{SC}$ the  spin of the resulting SC state at the 
impurity site.
The quantities ${\cal C}_2$ are the $SU(N)$ generalization of the 
$SU(2)$ eigenvalues, $S(S+1)$, and can be readily evaluated
(for details, see Ref. \cite{ljb}).

The ground state corresponds to having $n_c=(N-q)$ electrons at the
origin, partially screening the
impurity. It can be written explicitely as the action of $(2S-1)$ bosonic 
operators on a singlet state. For instance the {\it highest spin} state can
be written as
\begin{equation}
|GS\rangle _{\{a\}aa}^{[2S-1]}=\frac{1}{\scriptstyle\sqrt{(2S-1)!}}%
(b_{a}^{\dagger })^{2S-1}
\left[ \frac{1}{\gamma }{\mathcal A}(b_{i_{1}}^{\dagger
}(\prod_{\alpha =i_{2}}^{i_{q}}f_{\alpha }^{\dagger })(\prod_{\beta
=i_{q+1}}^{i_{N}}c_{\beta }^{\dagger }))\right]|0\rangle~ 
\end{equation}%
with $\gamma \equiv \sqrt{%
(2S+N-1)C_{N-1}^{q-1}}$.
Here, $c^\dagger_{\alpha} \equiv c^\dagger_\alpha(0)$.
We denote the ground state energy by $E_0$.

The effect of the kinetic term in Eq. \ref{ham1} is, to lowest order in
perturbation theory, to mix the ground state with excited states where the 
number of electrons changes by one. 
There are three such states, which we
denote by $|GS+1\rangle^S$, $|GS+1\rangle^A$ and $|GS-1\rangle$. 
The labels S(A) indicate that the additional electron is coupled
symmetrically(antisymmetrically) to the ground state. 
The states are
readily obtained by deriving the relevant SU(N) Clebsch-Gordan coefficients
\cite{ljb}.

\subsection{Stability of the strong coupling fixed point}

In order to better understand the low-energy physics of the system,
we should consider the finite Kondo coupling, allowing virtual hopping from
and to the impurity site. These processes generate interactions between the
composite at site 0 and the conduction electrons on neighboring sites, that 
can be treated as perturbations of the SC fixed point. 
The energy shifts due to the perturbation can be reproduced by introducing
an interaction between the spin at the impurity site and the spin
of the electrons on the neighboring site, with effective coupling $J_{ef\!f}$.
Applying 
an analysis similar to that of Nozi\`eres and Blandin \cite{nb}
to the nature of the 
excitations, we can argue whether or not the SC fixed point 
remains stable once virtual hopping is allowed. 

Thus, if the coupling 
between the effective spin at the impurity site and that of the electrons
on site 1 is ferromagnetic we know, from the scaling analysis at weak 
coupling, that the perturbation is irrelevant, and the low energy physics
is described by a SC fixed point. That is, an underscreened, 
effective
impurity weakly coupled to a gas of free electrons with a phase shift 
indicating that there are already $(N-q)$ electrons screening the original 
impurity.

If, on the contrary, the effective coupling is antiferromagnetic, the 
perturbation
is relevant, the SC fixed point is unstable and the 
low-energy physics of the model corresponds to some intermediate 
coupling fixed point,
to be identified. 

We explicitely calculate the effects of hopping on the
SC fixed point to the lowest order in perturbation theory,
that is, second order in $t$. We consider
the case with an arbitrary number $n_d$ of conduction electrons in site 1
generalizing the case $n_d=1$ considered in ref. \cite{cpt2}. 

\begin{figure}[ht]
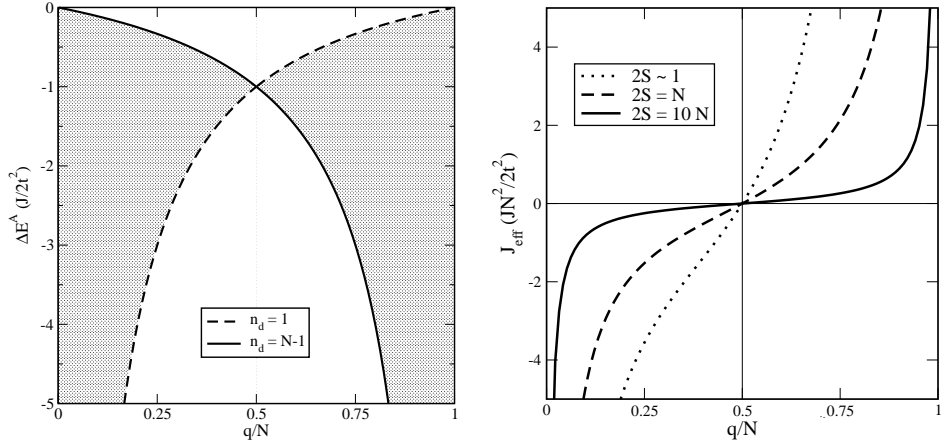

\centerline{
\begin{tabular}{c@{\hspace{1pc}}c}
\includegraphics[width=6.0cm]{e0d.eps} &
\includegraphics[width=6.0cm]{ejc.eps} 
\end{tabular}}
\caption {
(Left) Leading term in the energy shift, $\Delta E_0^A \sim \Delta E_0^S$, as a function 
of $q/N$, 
for $1<n_d<N-1$ (shaded region), and in
the limiting cases  $n_d/N \ll 1$ (dashed line), and  $n_d/N \approx 1$ 
(solid line). Notice that the
value at $q/N=1/2$ is equal to $-2t^2/J$, for any $n_d$. 
(Right) Energy difference, ($\Delta E_0^S-\Delta E_0^A$), as a function of
$q/N$, for different values of $2S$ in the large-N limit.
}
\label{qpure}
\end{figure}

Adding $n_d$ electrons on site 1
leads to two different states, that we will call symmetric (S) and 
antisymmetric (A). These states, degenerate in the SC limit,
acquire energy shifts, $\Delta E^A_0$ and $\Delta E^S_0$
due to the perturbation given, in the large-N limit, by 
\begin{eqnarray}
\Delta E_0^A = -\left(\frac{2t^2}{J}\right)\left[ 
\left(\frac{N-q}{q}\right)
-\frac{n_d}{N}\left(\frac{N(N-2q)}{q(N-q)}\right)\right]~,
\label{rdea}
\end{eqnarray}
\begin{eqnarray}
\Delta E_0^S-\Delta E_0^A &=& 
-\left(\frac{2S+n_d-1}{2S+N-q}\right)\left(\frac{2t^2}{JN}\right)
\left(\frac{N(N-2q)}{q(N-q)}\right)~,~
\label{resu} \\ &=& \frac{J_{ef\!f}}{2}(2S+n_d-1)~. \nonumber 
\end{eqnarray}
Notice that the behavior of both Eq. \ref{rdea} and Eq. \ref{resu} are
controlled by the same factor. 
This result has the immediate following physical consequence.
The change of sign of $J_{ef\!f}$ (Fig. \ref{qpure}, Right) 
-and hence of the stability of the SC
fixed point- is directly connected to the change in the behavior of 
$\Delta E_0^A \sim \Delta E_0^S$ with $n_d$.
In particular, when $\Delta E_0^S=\Delta E_0^A$,
$\Delta E_0^A=-(2t^2/J)$ independently \footnote{This is true for
arbitrary values of $N$, and $2S$.} of $n_d$, $q$ or $2S$. 
In the regime where the SC fixed point is stable, 
$q/N < 1/2$, $J_{ef\!f}<0$, the lowest energy corresponds to $n_d= 1$, whereas 
for $q/N > 1/2$, $J_{ef\!f}>0$, the energy expressed in Eq. (\ref{rdea}) is
minimized for $n_d=(N-1)$ (Fig. \ref{qpure}, Left).
 This is precisely the mechanism behind
the two-stage quenching. The accumulation of electrons on site 1 is not related to $J_{ef\!f}$ which is
independent of $n_d$, but results from the dependence of $\Delta E_0^A \sim \Delta E_0^S$ with $n_d$.

We finish by making some remarks on the physical properties of the model in the different
regimes. As is common to all models with an antiferromagnetic Kondo coupling, there will be 
a crossover from weak coupling above a given Kondo scale, $T_K$, to a low-energy regime.
When the SC fixed point is stable, we should expect for $T\ll T_K$ a weak 
coupling of the effective impurity at
site-0 with the rest of the electrons. The physical properties at low temperature are
controlled by the degeneracy of the effective impurity, $d([2S-1])=C^{N-1}_{N+2S-2}$.
Thus, we should expect a residual entropy ${\cal S}^i \sim \ln C^{N-1}_{N+2S-2}$ and a
Curie susceptibility, $\chi^i \sim  C^{N-1}_{N+2S-2}/T$, with logarithmic corrections \cite{jaz,pgks}. 
This is the result that we would
expect for a purely symmetric impurity. 
Contrary to the purely bosonic case,
only $(N-q)$ electrons are allowed at the origin in the SC limit, instead of
$(N-1)$. Thus, we would expect to find different results for quantities that involve the
scattering phase shift of electrons off the effective impurity.

In the $q>N/2$, we do not have access to the intermediate coupling fixed point that determines the low-energy
behavior. Nevertheless, it is reasonable to think that there would be a magnetic contribution
to the entropy, and a Curie-like contribution to the susceptibility, since the impurity
remains unscreened. This behavior is different from that of the multichannel
Kondo model, which is also characterized by an intermediate coupling fixed
point, but where the impurity magnetic degrees of freedom are completely
quenched. The degeneracy of the true ground state is an open question, but we
can assume that the entropy will be smaller than that of the SC fixed point \cite{aflw1}.
It is in the scattering properties that we might be able to see the anomalous
features of this new fixed point more clearly.

\section{Conclusions}
We have studied a Kondo model where the spin of the impurity has mixed 
symmetry, as a way of incorporating the phenomena of local moment screening 
and 
magnetic correlation that is observed in some heavy fermion compounds.
Such model is naturally realized by extending the spin symmetry to
$SU(N)$, and it displays a rich phase diagram. We find that as long as
the bosonic component of spin is of order $N$, there is a transition 
around the point where the fermionic 
component of the impurity is $q=N/2$. At this particular point, the energy 
shift
is, to lowest order in perturbation theory around the
SC fixed point, equal to $-2t^2/J$, independent of the
impurity parameters, $q$, $S$ and $N$. When $q<N/2$, the low-energy 
physics corresponds to the SC fixed point. For $q>N/2$
the SC fixed point is unstable and anomalous behavior is
expected, in particular, the two-stage quenching effect. This phase
diagram is not accidental, but is due to the relation of the 
effective impurity site in the SC regime to the conduction
electrons in neighboring sites, as our study of the dependence of the
energy shifts on
$n_d$ reveals. If $q<N/2$, the energy is minimized when the dressed
impurity repeals the electrons on the next site. That is,
when $n_d=1$. At $q=N/2$, the energy shift is also independent of $n_d$.
Finally, the lowest energy shift for $q>N/q$ corresponds to a maximal
$n_d$, indicating the accumulation of electrons leading to the two-stage
Kondo quenching.

\vspace{0.2in}

\begin{acknowledgements}

We would like to thank V. Zlatic and A. Hewson for organizing this
very interesting meeting. We are most grateful to N. Andrei for 
continuous encouragement and discussions. We also thank S. Burdin,
P. Coleman, Ph. Nozi\`eres, and C. P\'epin for helpful discussions.

\end{acknowledgements}

\vspace{0.2in}

$^*$  Also at the Centre National de la Recherche Scientifique (CNRS). 

\vfill\eject

\end{document}